\documentclass[aps,prl,twocolumn,showpacs,amsmath,amssymb]{revtex4}
\usepackage{graphicx} 
\usepackage{dcolumn} 
\usepackage{bm} 
\begin{document}

\title{Ultrafast Interference Imaging of Air in Splashing Dynamics} 

\author{Michelle M.\ Driscoll}
\author{Sidney R.\ Nagel} 
\affiliation{ The James Franck Institute and Department of Physics, The University of Chicago }%

\date{\today}

\begin{abstract}
A drop impacting a solid surface with sufficient velocity will emit many small droplets creating a splash.  However, splashing is completely suppressed if the surrounding gas pressure is lowered.  The mechanism by which the gas affects splashing remains unknown. We use high-speed interference imaging to measure the air beneath all regions of a spreading viscous drop as well as optical absorption to measure the drop thickness.   Although an initial air bubble is created on impact, no significant air layer persists until the time a splash is created.  This suggests that splashing in our experimentally accessible range of viscosities is initiated at the edge of the drop as it encroaches into the surrounding gas.
\end{abstract}

\pacs{47.20Gv,47.20.Ma,47.55.D-}

\maketitle

When a liquid drop hits a surface, it may rebound~\cite{quere_bounce}, spread smoothly, or shatter violently in a splash, as first photographed by Worthington~\cite{worth_nat}.  Controlling whether a liquid splashes has important consequences in many applications, including fuel dispersion in the automative industry, splat formation in coating technologies, and pesticide application in agriculture~\cite{Yarin,Rein_rev}.  Liquid and surface properties obviously influence impact dynamics~\cite{Mundo,Range,Stow,basaran,lohse}; it is quite counter-intuitive however that lowering the ambient pressure eliminates splashing altogether~\cite{LeiPRL,LeiPRE,2010_nagelPRE,Yoon_2010}.  To measure transient air-layer dynamics, we develop a technique that combines the high spatial precision of interferometry (nm scale) with high time resolution ($15 \mu s$).  

At impact, a small amount of air is trapped beneath the falling drop, creating a bubble~\cite{Siggi_fingering,Siggi_center,Chandra,vanDam,Hicks}.  Recent theoretical work has suggested that this air pocket is linked to splashing dynamics~\cite{Mandre,BrennerJFM,Josserand_jet}.  In a sufficiently viscous liquid, splashing occurs at late times, several tenths of millisecond after impact~\cite{LeiPRE,2010_nagelPRE}.  This temporal separation between impact and splashing creates an ideal system to test whether the initial air pocket influences the later-time splashing dynamics.  Using our interference technique, we find the initial air cavity dynamics to be consistent with theoretical predictions~\cite{BrennerJFM}.  However, we find no significant air layer that persists beneath a spreading drop until the time of thin-sheet ejection --- a necessary precursor to splashing in high-viscosity liquids~\cite{2010_nagelPRE}.  Thus, an underlying air layer is not responsible for splashing in this high-viscosity regime.
 
\begin{figure*}[htbp] 
\includegraphics[width=6.5in]{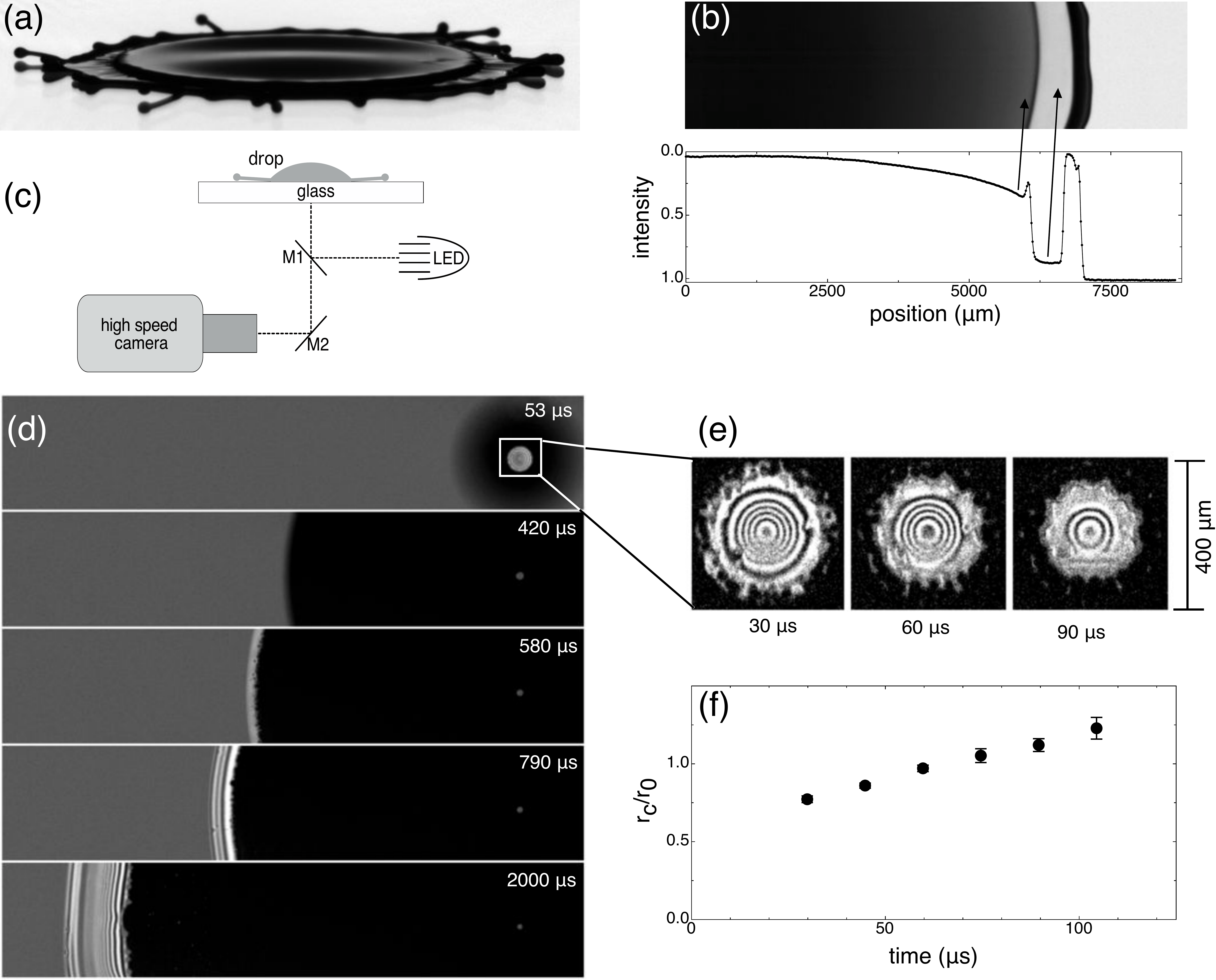} 
\caption{Images of a spreading drop after impact onto dry glass.  $\textbf{(a)}$ Side view showing thin-sheet ejection and splashing for a drop with $\nu$=15 cSt, and $u_0$=$3.5 $m/s.   $\textbf{(b)}$ Bottom view and intensity profile of a dyed $\nu$=9 cSt,$u_0$=$3.3$m/s drop used to meaure liquid thickness: the ejected sheet is approximately 1/10 the thickness of the lamella.  Arrows indicate measured regions.  $\textbf{(c)}$ Schematic for reflected-light interference: a half-silvered mirror (M1) directs light from a monochromatic LED onto the impacting drop from below; a standard mirror (M2) redirects reflected light into a high-speed camera.    $\textbf{(d)}$ Example time-series interference image of an impacting drop with $\nu$=21 cSt and $u_0$=$3.5 $m/s.  The lamella is uniform in intensity, indicating an optically flat region.  The thin sheet is ejected in the third panel.  $\textbf{(e)}$  Magnified images of initial air cavity.  This initial air cavity closes completely well before the thin sheet is ejected.  $\textbf{(f)}$ Radius of curvature of cavity normalized by drop radius, $r_c/r_0$ vs.\ time showing the cavity is slender ($r_c \sim r_0$), and gently flattens in time. Complete cavity closure occurs $\sim$150 $\mu s$ after impact, well before splashing occurs.} 
\label{splash}
\end{figure*}

The drops used in this study were mixtures of water and glycerol, with kinematic viscosities, $\nu$, between $9$ and $58$ cSt. Over this range there is minimal variation in surface tension, $\gamma$ ($65$-$67$ dyn/cm) and density, $\rho$ ($1.1$-$1.2$ g/cm$^3$).  Drops of uniform radius, $r_0$=$2.05$ mm, were created using a syringe pump. For each impact, we used a fresh glass substrate (Fisherbrand coverslip) to prevent surface contamination. The impact velocity, $u_0$, was varied between $1.5$ and $4.1$ m/s by releasing drops from various heights within an acrylic tube that could be evacuated to varying ambient pressures, $P$, between $2$ and $102$ kPa.  Using a high-speed camera (Phantom v12, Vision Research), we imaged drop impacts as shown in Fig.\ $1a$. To determine the thickness of the liquid as a function of position and time, we measured the local optical absorption of a spreading drop of colored liquid (Brilliant Blue G dye in a $\nu$=$9$ cSst glycerol/water solution) as shown in Fig.\ $1b$; we converted the transmitted light intensity to liquid thickness by calibrating with a liquid wedge of known proportions. By modifying our setup as shown in Fig.\ 1$c$, we also measured the thickness of any air layer underneath the spreading liquid using interferometric high-speed imaging at speeds up to $67,000$ frames/second.  We used a monochromatic LED ($\lambda$=$660$ nm) with a small coherence length, $\sim10 \mu m$, as a light source, so that there would be no interference between the two sides of the glass substrate. Adding a small amount of dye to our liquid greatly minimized the reflected light from the upper liquid surface and eliminated any interference generated within the liquid itself.  

Fig.\ $1a$ shows a $\nu$=$15$ cSt drop at atmospheric pressure: after spreading smoothly as a thick lamella for $\sim0.4$ms, it ejects a thin liquid sheet that subsequently disintegrates into smaller droplets --- the splash.  Using optical absorption, Fig.\ $1b$, we find the lamella edge to have a thickness of $106 \pm 4 \mu m$, while the ejected sheet is ten times thinner, only $10 \pm 2 \mu m$ thick. This jump in thickness occurs over a lateral extent of only $\sim300 \mu m$. 

There are many distinct splashing regimes that display different scalings and even qualitatively different behavior~\cite{Yarin,deegan}.  For example, below $\sim3$ cSt, sheet (i.e., corona) formation occurs within a few $\mu s$ of impact, while above $3$ cSt sheet ejection is delayed~\cite{LeiPRE}.    While it is not at all clear that the instability is the same across all splashing regimes, it is nevertheless established that lowering the air pressure eliminates splashing in all cases~\cite{LeiPRL,LeiPRE}.  The higher viscosity liquids used in this study allow for a large separation in time and space between the initial air layer entrapment and the creation of a splash.  This allows us to test directly whether the initially trapped air layer persists to longer times to influence the splashing dynamics.

Our interference technique determines the air-layer thickness beneath the drop as it spreads.  Fig.\ 1$d$ shows an example interference image at several different times after impact: $(i)$ just after impact, there is a small air cavity (panel 1) $(ii)$ surrounding this tiny region, the spreading lamella is uniformly black indicating an optically flat surface (panel 2), and $(iii)$ underneath the ejected thin sheet (panels 3-5), interference fringes are widely spaced indicating a very shallow slope.  

The small cavity of air trapped under the impacting drop (Fig.\ $1e$) has been shown to be present under varying conditions,~\cite{Siggi_fingering,Siggi_center,Chandra,vanDam,Hicks} including above and below the splashing threshold~\cite{2010_nagelPRE}.  By measuring the interference fringes that are clearly observed inside the cavity, we can directly measure the cavity curvature as a function of time.  The air cavity is quite flat --- the radius of curvature at the top of the cavity, $r_c$, is comparable to  $r_0$. At impact, the overpressure at the edge of the cavity is predicted to be higher than at its center~\cite{BrennerJFM}. This suggests $r_c/r_0$ should increase in time as the overpressure causes the cavity to flatten; this is consistent with the data shown in  Fig.\ $1f$.  

\begin{figure}[htbp] 
\centering 
\includegraphics[width=3.1in]{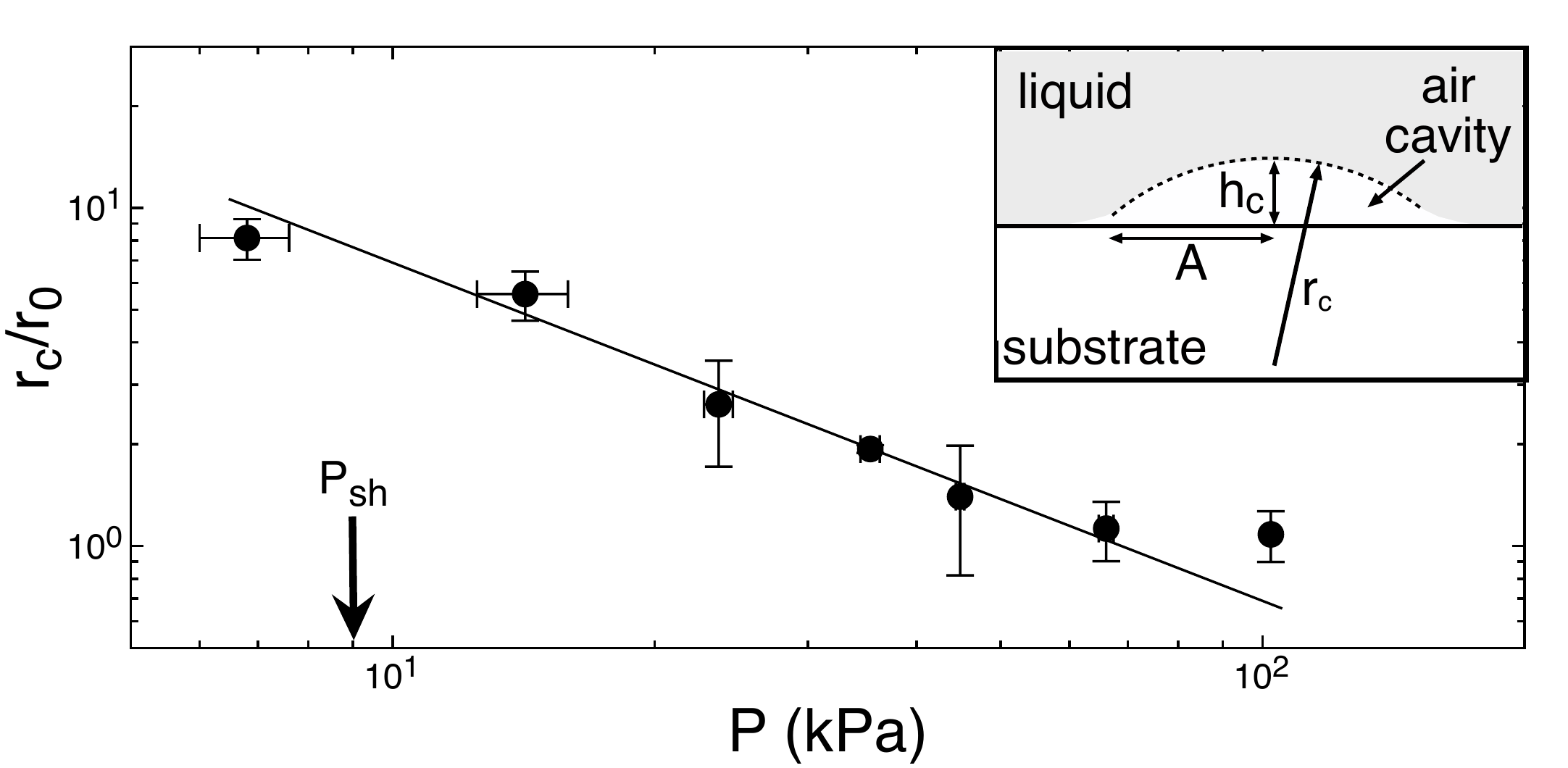} 
\caption{The radius of curvature of the entrapped air cavity normalized to drop radius, $r_c/r_0$, vs. ambient pressure $P$.  Inset schematic defines cavity height, $h_c$,  radius of curvature, $r_c$, and lateral cavity radius, $A$. As $P$ decreases, $r_c/r_0$ increases and the cavity becomes flatter. The line shows the best fit to $r_c/r_0 \propto P^{-1}$.}
\label{bubble}
\end{figure}

\begin{figure*}[htbp] 
\centering 
\includegraphics[width=6.75in]{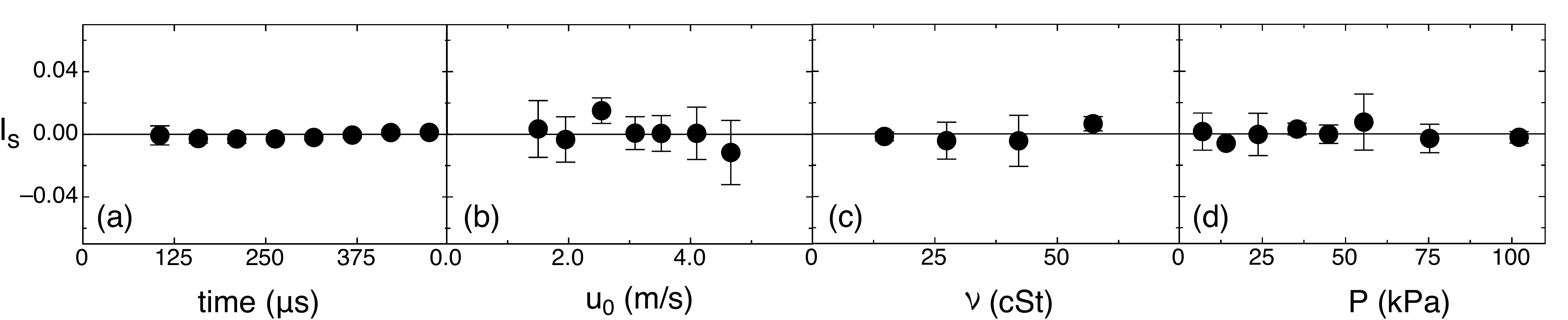} 
\caption{Interference signal, $I_s$, used to determine air-gap thickness, $h$, underneath the spreading lamella.  (a) $I_s$ vs.\ time, from the time the cavity collapses until just before sheet ejection.  (b-d) show $I_s$ measured one frame ($\sim 50 \mu s$) before sheet ejection as a function of (b) impact velocity, (c) viscosity, and (d) pressure.  These measurements place an upper bound on $h$ of 3.8 nm.  The data is consistent with the lamella making direct contact with the substrate as it expands.   Error bars represent drop to drop fluctuations.} 
\label{lamellaplots}
\end{figure*}

The air cavity persists even as the ambient pressure is decreased below the threshold value for sheet ejection, $P_{\mbox{\tiny{sh}}}$ \cite{2010_nagelPRE}.  However, the shape of the cavity strongly depends on pressure.  As shown in Fig.\ $2$, the cavity flattens dramatically with decreasing pressure: $r_c/r_0 \propto P^{-1}$.  To compare our curvature measurements with the theoretical prediction~\cite{BrennerJFM} that cavity height, $h_c \propto P$, we approximate the cavity as a thin spherical cap: 
\begin{equation}
h_c = \frac{A^2}{2r_c}, 
\end{equation}
where $2A$ is the lateral extent of the cavity (see Fig.\ $2$ inset).  Our measurement that $r_c/r_o \propto P^{-1}$ thus corroborates the prediction that $h_c \propto P$. We further note there is no transition in $r_c/r_0$ at $P_{\mbox{\tiny{sh}}}$, emphasizing that bubble entrapment does not appear to be related to the instability producing the thin sheet. Moreover, we note that the cavity closes into a bubble well before sheet ejection occurs.  We therefore conclude that the cavity dynamics is isolated in time and space from the edge of the spreading lamella where thin-sheet ejection and splashing take place.

Underneath the lamella, outside the bright spot created by the central entrapped bubble, our images are always nearly uniform and dark, see second panel of Fig.\ $1d$.  By carefully measuring the intensity in this region, we can constrain the height of any possible air film underneath the drop. 

For a liquid layer separated from a glass substrate by an air gap of height $h$, the total electric field produced by multiple reflections from the two interfaces is: 
\begin{eqnarray}
E_{L}(h) &=& \tfrac{n_g-1}{n_g+1} E_0 +\\  \nonumber
&&\tfrac{4 n_g (1-n_l)}{(1+n_g)^2(1+n_l)}E_0\sum_{k=1}^{\infty}\left[ \tfrac{(1 - n_g)(1-n_l)}{(1+n_g)(1+n_l)}\right]^{k-1} e^{i\delta k} ,
\label{sum}
\end{eqnarray}
where $n_g$ = 1.52 and $n_l$ = 1.44 + 0.0032$i$ are respectively the glass and liquid complex indicies of refraction.  The optical path length is given by $\delta = \frac{2\pi(2h)}{\lambda}$. The term $e^{i\delta k}$ in eqn.\ $2$ accounts for the phase shift caused by the path-length difference $2hk$ after the $k^{th}$ reflection from the lamella. Only the first few terms in eqn.\ $(2)$ are large enough to contribute significantly to the total electric field. Thus, the finite coherence length of the LED does not influence this calculation.  The total intensity is then $I_{L}(h)=|E_{L}(h)|^2+I_b$, where $I_b$ is an unknown background intensity from stray light and incoherent reflections.

To obtain the air gap thickness, $h$, we must eliminate the unknown quantities $|E_0|^2$ and $I_b$ from our expression for $I_{L}$. To do this, we measure $(i)$ the intensity under the lamella long after spreading has finished so that the liquid can be assumed to be in contact with the substrate,
\begin{equation}
I_c=\left|\frac{n_l-n_g}{n_l+n_g}E_0\right|^2+I_b,
\end{equation}
and $(ii)$ the intensity due to reflected light from only the substrate: \begin{equation}
I_1=\left|\frac{1-n_g}{1+n_g}E_0\right|^2+I_b.
\end{equation}
We can determine $h$ by computing:
\begin{equation}
I_s(h) =  \frac{I_L (h) - I_c}{I_1 - I_c}.
\label{signal}
\end{equation}

Fig. \ref{lamellaplots}$a$ shows $I_s$ from the time the cavity collapses into a bubble until just before sheet ejection. Fig.\ $3b$-$d$ shows $I_s$ measured just before the instant of sheet ejection vs.\ $u_0$, $\nu$, $P$. In all cases, the data remain essentially constant and within error of zero. The error bars in Fig.\ \ref{lamellaplots} represent drop-to-drop fluctuations.  
The distribution of all measurements of $I_s$ has a mean $0.0004$ with a standard deviation $0.0054$. This is comparable to the noise in the camera between adjacent frames when filming a still drop.   

These measurements of $I_s$ are consistent with the liquid being in direct contact with the substrate; this data places  an upper bound of 3.8 nm (using eqn.\ \ref{signal}) on the thickness of any possible air layer beneath the spreading lamella for all of the parameter space sampled.  An air layer of this thickness would be highly unstable and it is difficult to conceive that it could persist over $0.4$ ms, i.e.\ until the moment of sheet ejection. All of the air trapped beneath the falling drop is enclosed into the small central bubble discussed above and does not influence the subsequent sheet ejection and splashing. 

However, once the thin sheet is ejected, it does  move over a layer of air, which is easily visualized with our interference technique, see the bottom three panels in Fig.\ $1d$.  The sheet is ejected at a very shallow angle, varying from $0.1^\circ$ to 0.$25^\circ$, and this angle is relatively insensitive to $\nu$ and $P$ but decreases with increasing $u_0$. 

It is highly anti-intuitive that the surrounding gas controls splashing in all viscosity regimes.  Our interference technique allows quantitative measurements of the air beneath a spreading drop as a function of position and time.  In the low-viscosity splashing regime, corona formation occurs very near to the entrapped air bubble, both spatially and temporally.  Techniques such as total internal reflection imaging have been used to explore low-viscosity impact dynamics~\cite{Kolinski}, confirming theoretical predictions ~\cite{BrennerJFM} of the initial, transient air film.  In higher viscosity fluids, we find these initial air-cavity dynamics are also in quantitative agreement those predictions. 

However, we find no trapped air beneath the spreading drop outside the small central bubble; there is no significant air film beneath the drop at the time of thin-sheet ejection. This suggests that, rather than an underlying air layer, gas flow at the edge of the spreading drop is responsible for destabilizing the liquid. This conclusion is consistent with previous splash experiments in the low-viscosity regime~\cite{LeiPRL}. In that case, the scaling with gas pressure and molecular weight suggests that the liquid front expanding into the surrounding air leads to liquid destabilization and splashing.  The results reported here for more viscous fluids suggest a similar instability due to  leading-edge gas flows.

We thank Michael Brenner, Taehun Lee, Shreyas Mandre, Cacey Stevens, Lei Xu and Wendy Zhang for many fruitful discussions.  This work was supported by NSF-MRSEC grant No. DMR-0820054 and NSF grant No. DMR-1105145.  Use of facilities of  the Keck Initiative for Ultrafast Imaging are gratefully acknowledged.

\bibliography{splashingbib}

\end{document}